 \documentclass[doublecol]{epl2} 
\usepackage{bm,amsmath}
\usepackage{color}
\usepackage{psfrag}
\usepackage{graphicx}
\usepackage{epsfig}
\usepackage{amssymb}
\usepackage{color}

\definecolor{dblue}{rgb}{0.1,0.1,0.44}
\definecolor{dgreen}{rgb}{0.2 ,0.54, 0.2}
\begin{document}
\title{Kink confinement in the antiferromagnetic XXZ spin-$1/2$ chain
in  a  weak staggered magnetic field} 
\shorttitle{Kink confinement in the XXZ spin chain} 
\author{Sergei B. Rutkevich}
\institute{}
\bibliographystyle{eplbib}
\abstract{
The Heisenberg XXZ spin-1/2 chain is considered in the massive antiferromagnetic regime
in the presence of a staggered
longitudinal magnetic field. 
The Hamiltonian of the model is characterised by the anisotropy parameter $\Delta<-1$ and by the magnetic field strength $h$.
At zero magnetic field, the model is exactly solvable.  In the thermodynamic limit, it has two
degenerate vacua and the kinks   (which are also called spinons) interpolating between these vacua,  as elementary excitations. 
Application of the staggered magnetic field breaks integrability of the  model and induces the
long-range attractive potential between two adjacent kinks leading to their confinement into the bound states. The energy spectra of the resulting two-kink bound states {are} perturbatively calculated in
the extreme anisotropic (Ising) limit $\Delta\to-\infty$ to the first order in the inverse anisotropy constant  $|\Delta|^{-1}$, 
and also for any $\Delta<-1$ to the first order in the weak magnetic field $h$. 
}

\maketitle

\section{Introduction}
The confinement phenomenon occurs when the constituents of a compound particle cannot be separated from each other and, therefore, cannot be observed directly. A prominent and important example in high-energy physics is the confinement of quarks in hadrons.
It is remarkable, that confinement can  also be realized  in such condensed matter systems, as  quantum 
quasi-one-dimensional   ferro- and anti-ferromagnets  \cite{Coldea10,Mor14,Gr15,Wang15,Bera17,Faur17}. The present
theoretical understanding  \cite{FZ06,Rut08a}  of the confinement in such systems originates from the Wu and McCoy scenario \cite{McCoy78}, in which the
two kinks are treated as quantum particles moving in the line and attracting one another with a linear potential proportional 
to the  external magnetic field. 

Very recently \cite{Bera17}, the magnetic excitations energy spectra
 in the quasi-one-dimensional spin-1/2 antiferromagnetic compound ${\rm {SrCo_2V_2O_8}}$ in the confinement regime 
have been studied by means of the inelastic neutron scattering. The experimentally observed energy spectra 
were interpreted in \cite{Bera17} in terms of the one-dimensional  
 XXZ spin-1/2 chain Hamiltonian. 
We write this Hamiltonian directly in the thermodynamic limit 
in a slightly different form using the more traditional parametrization (see, {\it e.g.}, eq. (1.1) in \cite{Jimbo95})
\begin{align}\nonumber
\mathcal{H}(\Delta,h)&=-\frac{1}{2}\sum_{j=-\infty}^\infty\!\!\left[\sigma_j^x\sigma_{j+1}^x+\sigma_j^y\sigma_{j+1}^y+
\Delta\left(
\sigma_j^z\sigma_{j+1}^z+1
\right)
\right]\\
&-h\sum_{j=-\infty}^\infty (-1)^j\sigma_j^z.\label{XXZH}
\end{align}
Here $\sigma_j^a$ are the Pauli matrices, $\alpha=x,y,z$,
$\Delta$ is the anisotropy parameter, 
$h$ is the strength of the staggered magnetic field, which mimics  \cite{Bera17} in the 1D Hamiltonian 
\eqref{XXZH} the weak interchain interaction in the 3D 
array of parallel spin chains  in the 3D ordered phase of the compound ${\rm {SrCo_2V_2O_8}}$.
The massive antiferromagnetic phase is realised at  $\Delta<-1$. The dynamical structure factors and the spectrum of magnetic excitations 
in the model \eqref{XXZH} were
numerically  studied in \cite{Bera17} in three different cases: (i) in the extreme anisotropic (Ising) limit $|\Delta|\gg 1$, (ii) close to the  isotropic point $\Delta\approx -1$, and (iii) for generic $\Delta\in (-\infty,-1)$.  The resulting energy spectra of the 
magnetic excitations were presented graphically in Figures 8 - 15, and in phenomenological formulas like 
(26), which contain  fitting parameters. 

The aim of the present paper is to  find analytic representations for the energy spectra of the two-kink
bound states in the whole Brillouin zone  in model \eqref{XXZH} expressed solely in terms of the Hamiltonian 
parameters $\Delta$ and $h$. The problem is solved perturbatively in two asymptotical regimes: in the 
extreme anisotropic limit $\Delta\to-\infty$, and  for any $\Delta<-1$ at small $h$.
 
\section{The Hamiltonian symmetries} 
The Hamiltonian \eqref{XXZH} commutes with the $z$-projection of the total spin
$$
S^z=\frac{1}{2}\sum_{j=-\infty}^\infty \sigma_j^z,
$$
and with the
 modified translation operator $\widetilde{T}_1={T}_1 U$, where  ${T}_1$ stands for the 
 unit translation, and $U=\otimes_{j\in\mathbb{Z}} \,\sigma_j^x$ is the global rotation by  $\pi$ 
 around the $x$-axis. 
For short, the operator $S^z$ will be called the "total spin" in the following.

\section{Ising limit}
In the  extreme anisotropic  limit $\Delta\to-\infty$, it is convenient to rescale the Hamiltonian \eqref{XXZH} 
 to the form
\begin{align}\nonumber
\mathcal{H}_I(\epsilon,h)=|\Delta|^{-1}\mathcal{H}(\Delta,h)\big|_{\Delta=-1/\epsilon}=
-\epsilon h \sum_{j=-\infty}^\infty (-1)^j\sigma_j^z \\
\label{HI}
+\sum_{j=-\infty}^\infty\left[\frac{1}{2}
\left(
\sigma_j^z\sigma_{j+1}^z+1
\right)-
\epsilon\,(\sigma_j^{+}\sigma_{j+1}^-+\sigma_j^-\sigma_{j+1}^+)
\right], 
\end{align}
where $\sigma_j^\pm=\frac{1}{2}(\sigma_j^x\pm i\sigma_j^y)$.
The ground states and the low-energy excitations 
of the Hamiltonian \eqref{HI}  can be effectively studied \cite{Jimbo95,Bera17} by means of the 
Rayleigh-Schr\"odinger perturbation theory in the small parameter $\epsilon$. We shall describe these 
straightforward   calculations to the first order in $\epsilon$ in order to gain insight into the 
nature of the two-kink bound states of the Hamiltonian \eqref{XXZH} in the case of a generic 
$\Delta<-1$.

The zero order  Hamiltonian  $\mathcal{H}_I(0,h)=\mathcal{H}_I(0,0)$ has two antiferromagnetic ground states
\begin{equation}
|\Phi_1\rangle=|\ldots\downarrow{\color{red}\
\underline{\stackrel{0}{\uparrow}}}\stackrel{1}{\downarrow}\uparrow\downarrow\ldots\rangle,\quad
|\Phi_2\rangle=|\ldots\uparrow{\color{red}\underline{\stackrel{0}{\downarrow}}}\stackrel{1}{\uparrow}\downarrow\uparrow\ldots\rangle, 
\end{equation}
which are degenerate in energy,  
$
\mathcal{H}_I(0,0)|\Phi_{1,2}\rangle=0.
$

The  localized one-kink states $|K_{\alpha\beta}(j)\rangle$  interpolate between the vacua 
$|\Phi_\alpha\rangle$ 
and $|\Phi_\beta\rangle$ to the left, and to the right, respectively, 
from  the bond connecting the sites $j$, $j+1$.  
In the state $|K_{\alpha\beta}(j)\rangle$, only two adjacent spins at the 
sites $j$ and $j+1$ have the 
same orientations.  
 {
 The one-kink states can be classified also by their spin $s=\pm1/2$.
 Namely
 $ S^z|K_{\alpha\beta}(j)\rangle=s\,|K_{\alpha\beta}(j)\rangle$,
where $s= 1/2$ if  the neighbouring spins  $j,j+1$ forming the kink are 
orientated "up", and $s= -1/2$ if  
the spins $j, j+1$   have the "down" orientation. 
For the given localized one-kink state $|K_{\alpha\beta}(j)\rangle$, let us denote by $\rho=0,1$ the parity of 
the kink location $j$, $\rho= j \,{\rm mod}\,2$. One can easily see, that the three 
discrete parameters $\rho,\alpha,s$ characterising the  localized  kink are not independent. 
For example, the  kink $|K_{12}(j)\rangle$ has the spin $s=1/2$ for even $j$, and $s=-1/2$ otherwise.
In the general case, the
relation between  parameters  $\rho,\alpha,s$ can be described by the function
$\rho(s,\alpha)=\frac{1}{2}\left[
1+(-1)^\alpha 2s
\right].$

In the topologically neutral sector, the lowest-energy excitations are 
the two-kink states.  }
The basis of localized  two-kink states  is formed by the vectors
$|K_{\alpha\beta}(j_1) K_{\beta\alpha}(j_2) \rangle$
where $j_1,j_2\in \mathbb{Z}$, $j_1<j_2$. 
 {
 These states can be classified by the total spin
 \begin{equation}
 S^z|K_{\alpha\beta}(j_1) K_{\beta\alpha}(j_2) \rangle=s\,|K_{\alpha\beta}(j_1) K_{\beta\alpha}(j_2) \rangle,
 \end{equation}
where $s=0,\pm1$, $s=s_1+s_2$, with $s_{1,2}$ denoting the spins 
of the individual kinks. One can easily see, that  $s=0$ for even $(j_2-j_1)$, and 
$s=\pm1$ for odd $(j_2-j_1)$. So, $|s|=\kappa$ where
$\kappa=(j_2-j_1)\,{\rm mod}\, 2$.
 }

Denote by $P_2$ the 
orthogonal projector onto the two-kink subspace $ \mathcal{L}_{2}$ spanned by 
 {
the basis $|K_{\alpha\beta}(j_1) K_{\beta\alpha}(j_2) \rangle$,}
and by  
 $ \mathcal{H}_2(\epsilon,h)=P_2 \mathcal{H}_I(\epsilon,h) P_2$  the restriction
of the Hamiltonian \eqref{HI} on $ \mathcal{L}_{2}$.
 {
 At $\epsilon=0$, all the two-kink states $|\Psi\rangle\in \mathcal{L}_{2}$ are characterized by the same energy,
$\mathcal{H}_I(0,h)|\Psi \rangle=2|\Psi\rangle$.  At $\epsilon>0$, 
this degeneracy is removed in the linear order in $\epsilon$. 
This allows one to restrict the first-order  analysis of the low-lying  excitation 
energy spectra of the Hamiltonian \eqref{HI}   to the subspace $ \mathcal{L}_{2}$.}

The reduced two-kink Hamiltonian  $\mathcal{H}_2(\epsilon,h)$ 
acts on the basis states of $ \mathcal{L}_{2}$ as follows,
\begin{align}\label{Ham2}
&\mathcal{H}_2(\epsilon,h)\,|K_{\alpha\beta}(j_1) K_{{\beta\alpha}}(j_2) \rangle=\\\nonumber
&\left[2+f_0 (j_2-j_1) \right]\,|K_{\alpha\beta}(j_1) K_{{\beta\alpha}}(j_2) \rangle\\
&-\epsilon\big\{
|K_{\alpha\beta}(j_1-2) K_{\beta\alpha}(j_2) \rangle+|K_{\alpha\beta}(j_1) K_{\beta\alpha}(j_2+2) \rangle+\nonumber\\\nonumber
&\left[|K_{\alpha\beta}(j_1+2) K_{\beta\alpha}(j_2) \rangle+|K_{\alpha\beta}(j_1) K_{\beta\alpha}(j_2-2) \rangle \right]\\\nonumber
&\times(1-\delta_{j_2-j_1,1})(1-\delta_{j_2-j_1,2})
\big\}.
\end{align}
Here $f_0=2h\epsilon$ is the "string tension", which determines the linear attractive potential acting between the two kinks.

At $h=0$, the reduced Hamiltonian  \eqref{Ham2} is diagonalised by the the two-kink states
$|K_{\alpha\beta}(p_1)K_{\beta\alpha}(p_2)\rangle_{s_1s_2}\in \mathcal{L}_2$
characterised by  the momenta $p_1,p_2\in \mathbb{R}/\pi \mathbb{Z}$, and the spins $s_1, s_2=\pm 1/2$
of the individual kinks,
\begin{eqnarray}
&&{\mathcal{H}_2(\epsilon,0)\,|K_{\alpha\beta}(p_1)K_{\beta\alpha}(p_2)\rangle_{s_1s_2}=}\\
&&{\left[\omega_0(p_1)+\omega_0(p_2)\right]
|K_{\alpha\beta}(p_1)K_{\beta\alpha}(p_2)\rangle_{s_1s_2},}\nonumber\\
\label{2kink}
&&|K_{\alpha\beta}(p_1)K_{\beta\alpha}(p_2)\rangle_{s_1s_2}=\\\nonumber
&&\sum_{m_1=
-\infty}^\infty\,\,\,
\sum_{m_2>m_1+(\rho_2-\rho_1)/2}^\infty\Big\{
\Big[
e^{i(p_1j_1+p_2 j_2)}+\\
&&S_{s}(p_1,p_2)e^{i(p_2j_1+p_1 j_2)}\Big]
|K_{\alpha\beta}(j_1)K_{\beta\alpha}(j_2)\rangle
\Big\}_{\begin{smallmatrix} j_1=2m_1-\rho_1\\ j_2=2m_2-\rho_2\end{smallmatrix} },
\nonumber
\end{eqnarray}
where $\rho_1=\rho(s_1,\alpha),\rho_2=\rho(s_2,\beta)$, 
$s=s_1+s_2=0,\pm1$,  $S_s(p_1,p_2)$ denotes the 
two-kink scattering amplitude
\begin{equation}
S_{s}(p_1,p_2)=-e^{i(p_1-p_2)|s|},
\end{equation}
{ and $\omega_0(p)$ is the kink dispersion law,
\begin{equation}
\omega_0(p)=1-2\epsilon \cos(2{p}).
\label{spctr}
\end{equation}}

The two-kink states \eqref{2kink} transform 
in a simple way under the action of the modified translation operator $\widetilde{T}_1$,
\begin{eqnarray}\label{mT}
\widetilde{T}_1|K_{\alpha\beta}(p_1)K_{\beta\alpha}(p_2)\rangle_{s_1s_2}=\\\nonumber
e^{i(p_1+p_2)}|K_{\alpha\beta}(p_1)K_{\beta\alpha}(p_2)\rangle_{-s_1,-s_2},
\end{eqnarray}
and 
satisfy the Faddeev-Zamolodchikov commutation relations, 
\begin{eqnarray}\label{FZ1}
&&|K_{\alpha\beta}(p_1)K_{\beta\alpha}(p_2)\rangle_{s_1s_2}=\\
&&S_{s_1+s_2}(p_1,p_2)
|K_{\alpha\beta}(p_2)K_{\beta\alpha}(p_1)\rangle_{s_1s_2}.\nonumber
\end{eqnarray}

At $h>0$,  the eigenvalue problem for the Hamiltonian $\mathcal{H}_2(\epsilon,h)$ 
can be  solved following the procedure described in \cite{Rut08a,Rut10C}. 
{
To these end, let us represent the eigenstate of the Hamiltonian $\mathcal{H}_2(\epsilon,h)$ 
in the form 
\begin{eqnarray} \nonumber
&&|\Psi (P,\kappa,\rho_1) \rangle =\sum_{m=-\infty}^\infty \sum_{r=1}^\infty \Big[e^{i P( j_1+r)}
\psi(r|P,\kappa,\rho_1)\\
&&\times |K_{12}(j_1) K_{21}(j_1+j) \rangle  \Big]_{j_1=2m-\rho_1, \,j=2r-\kappa},\label{eigv}
\end{eqnarray}
where $P\in\mathbb{R}/(\pi \mathbb{Z})$ is the total quasimomentum 
of two kinks, and the discrete parameters $\kappa$ and  $\rho_1$
take the values $0,1$. As it was explained earlier, the parameter $\kappa$ equals to 
the the absolute value of the total spin of the two-kink state, $\kappa=|s|$. 
The eigenstate problem $\mathcal{H}_2(\epsilon,h)|\Psi (P,\kappa,\rho_1) \rangle=E(P,\kappa)
|\Psi (P,\kappa,\rho_1) \rangle$ reduces due to \eqref{Ham2}
}
to the discrete Sturm-Liouville problem with the linear potential in the half-line 
for the wave function $\psi(r)$,
\begin{eqnarray}\label{BS0}
&&[2+f_0\,(2r-\kappa)-E(P,\kappa)]\psi(r)-\\\nonumber
&&2\epsilon \cos P\,\,\, [\psi(r+1)+\psi(r-1)]=0,
\end{eqnarray}
where $r=1,2,\ldots$, and the Dirichlet boundary  condition 
$ \psi(0)=0$ is imposed. Note, that the distance between the two kinks is $(2r-\kappa)$.

Exploiting the  equality
$$
J_{\nu+1}(Z)+J_{\nu-1}(Z)=\frac{2\nu}{Z}\,J_\nu(Z),
$$
one can  immediately  {write} down \cite{Gal85} the explicit solution of the  discrete 
Sturm-Liouville problem \eqref{BS0} in terms of the Bessel function $J_\nu(Z)$.
The resulting energy spectrum  reads 
\begin{equation}\label{La}
E_n(P,\kappa)=  2-2 \epsilon\, h\,[\kappa+2\nu_n(P)],
\end{equation} 
 where $\nu_n(P)$ are the solutions of the equation
 $$
J_{\nu_n(P)}(h^{-1}\cos P)=0, 
$$
  with $n=1,2,\ldots$.

Equation \eqref{La} determines,  for arbitrary $h>0$, the exact small-$\epsilon$ 
asymptotics for the energy spectrum of the two-kink bound states for the 
Hamiltonian \eqref{HI}  to the first order in $\epsilon$.
A very similar energy spectrum was found in \cite{Rut08a}  [see equations (54), (59), (60) therein],
where the kink confinement in the quantum Ising spin-chain  was studied.

In the case of small $h\to+0$, two asymptotical expansions can be obtained  \cite{Rut08a} from  
\eqref{La}, using the well-known properties of the Bessel function.
For not very large $n=1,2,3,\ldots$, the {\it low-energy expansion} in the fractional powers of $h$ holds,
\begin{align}\nonumber
E_n(P,\kappa)=2-4\epsilon \cos P +4\epsilon \left(\frac{\cos P}{2}\right)^{1/3} h^{2/3}\mathrm{z}_n\\
-2 \epsilon \kappa h+O(h^{4/3}),\label{spAi}
\end{align}
where $-\mathrm{z}_n$ are the zeroes of the Airy function. 
 In the case of large $n\gg 1$, one can use instead the {\it semi-classical expansion}
\begin{equation}\label{SC}
E_n(P,\kappa)=2-2\epsilon h \kappa-4\epsilon \cos P \, \cos(2 p_a(P)).
\end{equation}
Here $p_a(P)$ is the solution of two equations
\begin{align}
2p_a \lambda_a  +\sin(2p_a)=\frac{\pi h}{\cos P}\left(n-\frac{1}{4}\right)+O(h^2),\\
\cos(2p_a)=-\lambda_a,\label{eqk}
\end{align} 
which determine also the parameter $\lambda_a$. 
{Derivation of the asymptotic formulas \eqref{spAi}, \eqref{SC} from the
exact energy spectrum \eqref{La}
almost literally reproduces the derivation of the similar small-$h$ 
asyptotics for the exact energy spectrum in the Toy model 1, 
which was described in much detail in Section 6.1 of [8]. 
}

It turns out, that the small-$h$ asymptotic representations \eqref{spAi}, \eqref{SC} for the 
two-kink bound state energy spectra can be also derived
by  means of a different, semi-heuristic approach, which was initially developed for the Ising field theory \cite{Rut05,FZ06},
and then applied to  the quantum Ising spin-chain model \cite{Rut08a},
and to the Potts Field Theory (PFT) \cite{RutP09}. 
The high accuracy of the analytical predictions obtained by 
this technique was confirmed later \cite{Tak14,Kor16} by 
direct numerical  calculations of the kink bound state energy spectra in the confinement regime for all three models
mentioned above. In the cases of the Ising field theory and the Ising spin chain model, the
semi-heuristic technique  reproduces the initial terms of the asymptotical expansions for the bound state energy spectra obtained by the more powerful  technique based on the Bethe-Salpeter equation. For the PFT, 
the meson mass spectra calculation by means of the latter method  is still lacking. 
In what follows, we shall describe the semi-heuristic  technique for the case of the 
extreme anisotropic limit $\Delta\to-\infty$ of the antiferromagnetic XXZ model, and then apply it to the general case 
$\Delta<-1$.

Let us treat the two kinks as  classical particles having the $z$-projections of the spin  $s_i=\pm1/2$,  $i=1,2$,   
moving along the line, and attracting one another with a linear potential.
Their Hamiltonian will be taken in the form 
\begin{align}\label{Hk}
H(x_1,x_2,p_1,p_2,s_1,s_2)=\omega_0(p_1)+\omega_0(p_2)+\\
{f}_0\left[|x_2-x_1|-{\kappa({s_1,s_2})}\right].\nonumber
\end{align}
Here $x_1,x_2\in \mathbb{R}$ are the kink spacial coordinates, 
$\kappa({s_1,s_2})=\delta_{s_1,s_2}$, and $\omega_0(p)$ is the kink dispersion law \eqref{spctr}.

After the canonical transformation
\begin{subequations}\label{Canon}
\begin{align}
X=\frac{x_1+x_2}{2}, \quad x=x_2-x_1,\\
P=p_1+p_2, \quad p=\frac{p_2-p_1}{2},
\end{align}
\end{subequations}
the Hamiltonian \eqref{Hk} takes the form
\begin{equation}\label{Hk1}
H(p,x|P)=\varepsilon(p|P)+f_0 |x|,
\end{equation}
where 
\begin{equation}\label{eps}
\varepsilon(p|P)=\omega_0(p+P/2)+\omega_0(p-P/2)-f_0 \kappa.
\end{equation}
In order to simplify notations, we have dropped  here the spin  arguments $s_1,s_2$,
{
as well as the argument $\kappa$ in the 
function   $\varepsilon(p|P,\kappa)$ defined by \eqref{eps}.}

The topology of the phase trajectories in the $(x,p)$-plane depends on the 
total energy $E$ of the two kinks, as it is clear from Figure \ref{kinE}.
 The phase trajectories are closed for
$\varepsilon(0{|}P)<{E}<\varepsilon(\pi/2|P)$. In this case, 
the solution of the canonical equations   describes the oscillatory motion of two kinks in the 
center of mass frame that drifts with a constant average velocity. 
\begin{figure}[tb]
\begin{center}
\includegraphics[width=.8\columnwidth]{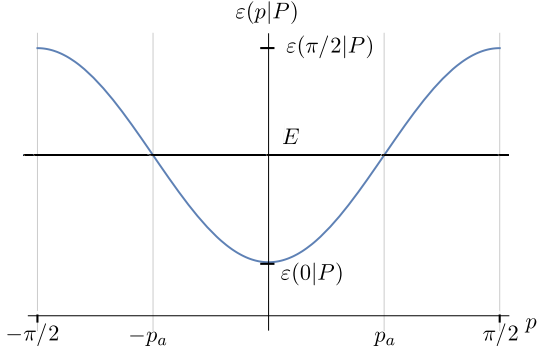}
\caption{\label{kinE} Kinetic energy \eqref{eps} of two kinks  as the function of $p$. 
For the energy $E$, the classically allowed region is $[-p_a,p_a]$.}
\label{kinE}
\end{center}
\end{figure}

There are two different ways to quantize  the model \eqref{Hk}. 
For small oscillations, the kinetic energy term  \eqref{eps} can be expanded 
to the second order in the momentum $p$, 
with subsequent replacement of the latter
by the  operator  $-i\partial_x$.  The resulting Schr\"odinger equation
can be reduced to the Airy equation  in the half-line $0<x<\infty$, which, together with 
the Dirichlet boundary condition at $x=0$, gives rise to
the energy spectrum \eqref{spAi}. 
For the high-amplitude oscillations with energies in the interval 
$\varepsilon(0{|}P)<E<\varepsilon(\pi/2{|}P)$,
the energy levels can be found  \cite{Rut08a} by means of the Bohr-Sommerfeld quantization rule. The resulting spectrum
reads 
\begin{equation}\label{WKB}
2 E_n(P,\kappa) p_a- \int_{-p_a}^{p_a} dp\, \varepsilon(p|P)=2\pi f_0 \,(n-1/4),
\end{equation}
with {$E_n(P,\kappa)=\varepsilon(p_a|P)$ and} $n\gg1$, which is equivalent to \eqref{SC}-\eqref{eqk}. 
{ Note, that the left-hand side of \eqref{WKB} is the Legendre transform of the integral $\int_{-p_a}^{p_a} \varepsilon(p|P)\,dp$
considered as a function of  the variable $p_a$.}
\section{General case $\Delta<-1$}  
Let us turn to the general case of the XXZ spin-chain model \eqref{XXZH} in the 
antiferromagnetic phase $\Delta<-1$.
We shall use the standard parametrisation   for the anisotropy constant $\Delta=-\cosh \gamma$,
and $q=\exp \gamma>1$.

At zero field $h=0$, the model considered on a finite chain 
is solvable by the Bethe Ansatz method \cite{Orb58}, see also \cite{Zv10,Dug15} for further references.
In the thermodynamic limit, it has two   
degenerate ground states $|\Phi_{\alpha}\rangle$, $\alpha=1,2$, showing a N\'eel-type order, 
$$
\langle\Phi_1| \sigma_j^z|\Phi_1\rangle=-\langle\Phi_2| \sigma_j^z|\Phi_2\rangle=(-1)^j\bar{\sigma},
$$
with  the spontaneous magnetization  \cite{Baxter1973,Baxter1976,Iz99}
\begin{equation}\label{sig}
\bar{\sigma}=\prod_{n=1}^\infty \left(
\frac{1-q^{-2 n}}{1+q^{-2 n}}
\right)^2.
\end{equation}

The lowest-energy excitations are topologically charged, being represented \cite{FT85,Jimbo95} by  the
kinks $|K_{\alpha\beta}(p)\rangle_s$ interpolating between the vacua $\alpha$ and $\beta$,
and characterised by the quasimomentum $p\in \mathbb{R}/(\pi \mathbb{Z})$, and by the 
$z$-projection of the spin $s=\pm1/2$. 
The dispersion relation of these excitations reads  \cite{McCoy73},
\begin{equation}\label{dispr}
\omega(p)=\frac{2K}{\pi}\,\sinh\gamma \,\sqrt{1-k^2 \cos^2 p} .
\end{equation}
Here $K$ and $K'$ are the complete elliptic integrals of modulus $k$ and $k'=\sqrt{1-k^2}$ respectively, such that $K'/K = \gamma/\pi$.
The dispersion  relation \eqref{dispr}  can be parametrized in terms of the Jacobi elliptic functions
\begin{eqnarray}\label{rap}
p(\lambda)=\frac{\pi}{2}-{\rm am}(2K \lambda/\pi,k),\\
\omega(\lambda)= \frac{2 K }{\pi}\,\sinh \gamma\,\, {\rm dn}(2K \lambda/\pi,k),
\end{eqnarray}
with the rapidity variable $\lambda\in[-\pi/2,\pi/2]$.

The number of kinks must be even in the topologically neutral sector. 
The two-kink basis states
$|K_{\alpha\beta}(p_1)K_{\beta\alpha}(p_2)\rangle_{s_1s_2}$, characterized by the 
quasimomenta $p_{1,2}$ and the spins $s_{1,2}$ of the individual kinks, 
 diagonalize the Hamiltonian $\mathcal{H}(\Delta,0)$ and the total spin $S^z$ with the eigenvalues
$\omega(p_1)+\omega(p_2)$, and $s_1+s_2$, respectively. The translation properties of these states 
are determined by equation  \eqref{mT}. In the extreme anisotropic limit $\gamma\to\infty$, the two-kink states
$|K_{\alpha\beta}(p_1)K_{\beta\alpha}(p_2)\rangle_{s_1s_2}$ reduce to \eqref{2kink}.

Let us define the following basis in the two-kink subspace with $S^z=0$,
\begin{align}\nonumber
&|K_{\alpha\beta}(p_1)K_{\beta\alpha}(p_2)\rangle_\pm\equiv
\frac{1}{\sqrt{2}}\Big(
|K_{\alpha\beta}(p_1)K_{\beta\alpha}(p_2)\rangle_{1/2,-1/2}\\
&\pm
|K_{\alpha\beta}(p_1)K_{\beta\alpha}(p_2)\rangle_{-1/2,1/2}
\Big).\label{baspm}
\end{align}
The  modified translation operator  becomes diagonal in this basis:
\begin{equation}\label{trmod}
 \widetilde{T}_1|K_{\alpha\beta}(p_1)K_{\beta\alpha}(p_2)\rangle_\pm=
\pm e^{i(p_1+p_2)}|K_{\alpha\beta}(p_1)K_{\beta\alpha}(p_2)\rangle_\pm.
\end{equation}
The two-kink scattering at $h=0$ can be described by the Faddeev-Zamolodchikov commutation relations:
\begin{align}\label{FZ2}
|K_{\alpha\beta}(p_1)K_{\beta\alpha}(p_2)\rangle_{ss}=w_0(p_1,p_2)
 |K_{\alpha\beta}(p_2)K_{\beta\alpha}(p_1)\rangle_{ss},\\
|K_{\alpha\beta}(p_1)K_{\beta\alpha}(p_2)\rangle_{\pm}=w_\pm(p_1,p_2)\nonumber 
|K_{\alpha\beta}(p_2)K_{\beta\alpha}(p_1)\rangle_\pm.
\end{align}
The three scattering amplitudes $w_\eta(p_1,p_2)$, with $\eta=0,\pm$, can be  parametrized by the 
rapidity variable, 
\begin{align}
&w_\eta(p_1,p_2)=\exp[-i\pi +i \theta_\eta(p_1,p_2)],\\
&\theta_\eta(p_1,p_2)=\Theta_\eta(\lambda_1-\lambda_2),\label{thet}\\
\label{Th0}
&\Theta_0(\lambda)=
- \lambda-\sum_{n=1}^\infty \frac{e^{-n\gamma}\sin(2\lambda n)}{n \cosh(n \gamma)},\\
&\Theta_\pm(\lambda)=\Theta_0(\lambda)+\chi_\pm (\lambda),\\
&\chi_+(\lambda)=-i \ln\left[- \frac{\sin(\lambda-i \gamma)/2)}{\sin(\lambda+i \gamma)/2)} \right],\\
&\chi_-(\lambda)=-i \ln\left[ \frac{\cos(\lambda-i \gamma)/2)}{\cos(\lambda+i \gamma)/2)} \right],
\end{align}
where $p_j=p(\lambda_j)$, $j=1,2$,  and $\Theta_\eta(\lambda)$ are the scattering phases.
The  scattering amplitude $w_0(p_1,p_2)$ was found by 
Zabrodin \cite{Zabr92},  and the whole two-kink scattering matrix was determined by 
Davies {\it et al.} \cite{Miwa93}. In the extreme anisotropic limit $\gamma\to\infty$,
the commutation relations \eqref{FZ2} reduce to \eqref{FZ1}.

The application of a staggered magnetic field $h>0$ breaks  the  integrability of the XXZ model and leads 
at $\Delta<-1$ to confinement of the kinks into the bound states. The natural way to study their 
energy spectrum is to apply some perturbative technique in small $h$
around the exact solution at $h=0$. The most systematic, but technically rather hard realization of this idea should
exploit the Bethe-Salpeter equation \cite{FonZam2003,FZ06}, together with the appropriate
form factor perturbative expansion \cite{Rut09,Rut17P}. Here we shall apply instead the more simple semi-heuristic 
approach, which was outlined above. Since the vacuum $|\Phi_2\rangle$ becomes metastable at
$h>0$, we shall concentrate in the following 
on the topological neutral sector  spanned by the basis states $|K_{12}(p_1)K_{21}(p_2)\rangle_{s_1s_2}$.

So,  let us consider  two interacting particles moving in the line, 
whose classical evolution is determined by the Hamiltonian 
\begin{equation}\label{Hk2}
H(x_1,x_2,p_1,p_2)=\omega(p_1)+\omega(p_2)+
{f}\,|x_2-x_1|.
\end{equation}
Now the particle kinetic energy $\omega(p)$ is  taken in the form \eqref{dispr}, 
and for the  string tension we use its value 
$ f=2 h \bar{\sigma}$ at $h\to+0$,
where the spontaneous magnetisation $\bar{\sigma}$ is given by \eqref{sig}. 
 Quantization of the periodical 
 motion of two particles in the center of mass  frame {should allow} one to determine the energy
 spectrum of the {two-kink bound states of model \eqref{XXZH}
at $h\to+0$.} 
\begin{figure}[tb]
\begin{center}
\includegraphics[width=.85\columnwidth]{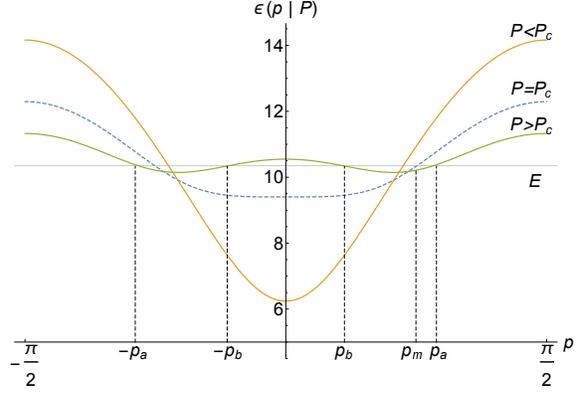}
\caption{\label{kinE2} Kinetic energy \eqref{eps1} of two kinks  at $\Delta=-5$ as the function of $p$
for three values of their  total momentum $P$: $P<P_c$, $P=P_c$, and $P>P_c$. 
For the energy $E$ in the latter case, 
the classically allowed regions are $[p_b,p_a]$ and $[-p_a,-p_b]$.}
\label{epsilon}
\end{center}
\end{figure}

Two new features, which modify the analysis,
{must} be taken into account.  
First, due to the different kink dispersion law  \eqref{dispr}, the profile of the effective kinetic energy $\varepsilon(p|P)$ in the 
center of mass frame 
\begin{equation}\label{eps1}
\varepsilon(p|P)=\omega(p+P/2)+\omega(p-P/2),
\end{equation}
now transforms with increasing total momentum $P$, as it is shown in Figures \ref{kinE2}.
At small  total momenta $P$, the kinetic energy $\epsilon(p|P)$ takes its minimal value at the origin $p=0$, 
and monotonically increases with $p$ at $0<p<\pi/2$. At large enough $P$, the kinetic energy becomes non-monotonic. 
It has a local maximum at
$p=0$, and two minima located at $p=\pm p_{m}(P)$.
The transition between these two regimes takes place at the critical  value   
$P_c(k')=\arccos\left(\frac{1-k'}
{1+k'}
\right)$ of the total momentum.
As the result,  the classical phase portrait of the two particle relative motion changes  at $P>P_c(k')$, which also 
affects the quantization of their dynamics. 

Fortunately, the kink dispersion law \eqref{dispr} in the antiferromagnetic XXZ-model
coincides up to a re-parametrization with the kink dispersion law  
in the Ising spin-$1/2$ chain in a  transverse magnetic field. This fact allows one to 
apply the results of the paper \cite{Rut08a}, in which the same semi-heuristic approach has been
used to calculate the two-kink energy spectrum in 
the latter model, together with the more systematic  method based on the Bethe-Salpeter equation. 

Then, in contrast to the Ising model, the kinks in the XXZ-model are not free particles, but strongly interact 
at small distances already at $h=0$. This short-range interaction leads to the nontrivial two-kink scattering, which 
must be properly taken into account. The problem of  kink confinement 
in the presence of a  nontrivial kink-kink scattering has been already studied  in the case of the PFT \cite{RutP09}. 
The dynamics of two kinks confined into a bound state in the semiclassical regime at $h\ll 1$ was described 
in \cite{RutP09} in two different ways. At large separations $|x_2-x_1|$, the two kinks were treated as classical 
particles  moving according to the canonical equations of motion. 
When the two kinks approach one another at some point $x_2=x_1$ 
having the momenta $p_1$ and $p_2$, they undergo  quantum scattering, which is described
by the Faddeev-Zamolodchikov commutation relations analogous to 
\eqref{FZ2}. As the result, the semiclassical energy spectrum
of the two-kink bound states  becomes explicitly dependent 
on the kink-kink scattering phases. Applying the same  strategy, we calculated the energy spectrum of the
two-kink bound states in the XXZ-model for any $\Delta<-1$ to the first order in $h\to+0$. Here we shall present the results only. 
The details of the calculations, which are to much extent 
similar to those described in papers \cite{Rut08a,RutP09}, will be published elsewhere. 

There are three spectral modes, which will be  distinguished by the 
parameter $\eta$ taking the "values" $0$ and $\pm$. The twofold degenerate mode with $\eta=0$ 
 corresponds to the kink bound states with $S^z=\pm 1$.
Two other modes  $\eta=\pm $   correspond to the 
 kink bound-states with $S^z=0$. 
 The wave functions of  such states  can be 
 expanded in the bases  \eqref{baspm},
which are diagonalize the scattering matrix at $h=0$. These two $S^z=0$ modes, which  
are degenerate in the Ising limit $\Delta\to-\infty$,
split at finite $\Delta<-1$ due to the difference in their two-kink scattering phases.
{ The lowest energy of all  three spectral modes has the mode with  $\eta=0$.}

\underline{At $|P|<P_c(k')$}, the initial terms of the low energy expansion take the form,
\begin{align}\nonumber
E_n(P,\eta)=2 \,\omega(P/2)+
f^{2/3}[\omega''(P/2)]^{1/3}{\text z}_n\\
+f\frac{\sinh\gamma}{\omega(P/2)}\partial_\lambda\Theta_\eta(\lambda)\big|_{\lambda=0}+O(f^{4/3}),
\label{lowE}
\end{align}
where $n=1,2,\ldots$, and $-{\text z}_n$ are the zeroes of the Airy function.
So, the shifts between the energy spectra of the three modes with different
$\eta$ are proportional to the magnetic field.

The leading term of the semiclassical expansion for the energy spectra of all three modes at $|P|<P_c(k')$ reads 
\begin{align}\label{EN}
&2 E_n(P,\eta)\,p_a-\int_{-p_a}^{p_a} \varepsilon(p|P)\,dp=\\\nonumber
&f\left[
2\pi \left(n-\frac{1}{4}\right)+\theta_\eta\left(\frac{P}{2}-p_a,\frac{P}{2}+p_a \right)
\right]+O(f^2),
\end{align}
where $n\gg 1$, $\theta_\eta(p_1,p_2)$ are the scattering phases  defined by 
\eqref{thet}, and  $p_a\in(0,\pi/2)$  is the solution of the equation
$E_n(P,\eta)=\varepsilon(p_a|P)  $.  
\begin{figure}[tb]
\begin{center}
\includegraphics[width=\columnwidth]{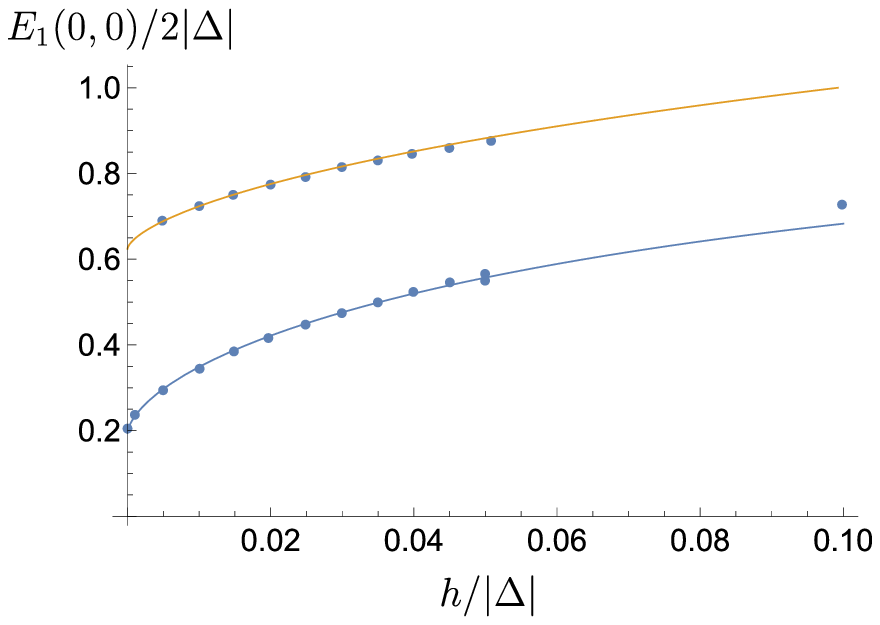}
\caption{The energy  of the first bound  state for the mode  $\eta=0$ at $P=0$ versus the magnetic field
calculated from \eqref{lowE} at $\Delta=-5$ (upper curve), and at $\Delta=-2$ (lower curve). 
The points display the results of numerical calculations on the finite chains  at the same values of parameters 
performed  
by Bera {\it et al.} and presented in Figure 11 of  \cite{Bera17}.}
\label{Fig:spectra}
\end{center}
\end{figure}

\underline{At $P_c(k')<|P|<\pi/2$}, the low-energy expansion takes
the form
\begin{align}
&E_n^{(1,2)}(P,\eta)=\varepsilon(p_m|P)+\nonumber
f^{2/3}\left[
\frac{\partial_p^2\varepsilon(p|P)\big|_{p=p_m}}{2}
\right]^{1/3}x_n^{(1,2)}\\
&+\frac{f}{2}\, \partial_p \,\theta_{\eta}(P/2-p,P/2+p)\big|_{p=p_m} 
+O(f^{4/3}),\label{En12}
\end{align}
where $p_m=\frac{1}{2}\arccos(\cos P/\cos P_c)$ is the 
location of the minimum of the kinetic energy $\varepsilon(p|P)$, and $-x_n^{(1)}=-\mathrm{z}_n$ and $-x_n^{(2)}=-\mathrm{z}_n'$ are the 
zeroes of the the Airy function and of its derivative, respectively, $Ai(-\mathrm{z}_n)=0$, $Ai'(-\mathrm{z}_n')=0$, $n=1,2,\ldots$. 
The semi-classical asymptotics at  $E_n(P,\eta)\in (\varepsilon(p_m|P),\varepsilon(0|P))$ modifies to the form
\begin{align}\nonumber
 &E_n(P,\eta)\,(p_a-p_b)-\int_{p_b}^{p_a} \varepsilon(p|P)\,dp=f\pi \left(n-\frac{1}{2}\right)\\\nonumber
&+\frac{f}{2}\left[
\theta_\eta\left(\frac{P}{2}+p_b,\frac{P}{2}-p_b \right)+\theta_\eta\left(\frac{P}{2}-p_a,\frac{P}{2}+p_a \right)
\right]\\\label{EN1}
&+O(f^2),
\end{align}
where $p_{a,b}\in(0,\pi/2)$ are the positive solutions of the equation
\begin{equation}
E_n(P,\eta)=\varepsilon(p_a|P)=\varepsilon(p_b|P), \quad p_b<p_a.
\end{equation}
For the energies in the interval $E_n(P,\eta)\in (\varepsilon(0|P),\varepsilon(\pi/2|P))$, the semiclassical spectrum is described
by equation \eqref{EN}.

\underline{At $|P|= P_c(k')$}, the Taylor expansion of the kinetic energy $\epsilon(p|P_c)=\epsilon(0|P_c)+\frac{p^4}{4!} \partial_p^4\varepsilon(p|P_c)\big|_{p=0}+\ldots$ 
does not contain the quadratic term. As the result, the low-energy expansion 
changes to the form
\begin{align}\nonumber 
&E_n(P_c,\eta)=2\omega(P_c/2)
+f^{4/5}\left[
\frac{\partial_p^4\varepsilon(p|P_c)\big|_{p=0}}{6}
\right]^{1/5}\, c_n+\\
&f\frac{\sinh\gamma}{\omega(P_c/2)}\partial_\lambda\Theta_\eta(\lambda)\big|_{\lambda=0}+O(f^{8/5}),\label{LE54}
\end{align}
where $c_n$ are the consecutive solutions of the equation (93) in \cite{Rut08a}. The numerical values of the
first three ones are  $c_1 = 1.787$, $c_2 = 3.544$, 
$c_3 = 5.086$. The low-energy expansion \eqref{LE54} holds for the energies slightly above the lower bound of the spectrum, 
$E\approx \varepsilon(0|P_c)$. For higher energies in the interval $\varepsilon(0|P_c)<E<\varepsilon(\pi/2|P_c)$, 
the semiclassical asymptotics  (\ref{EN})  can be used.

\section{Conclusions} 
The energy spectrum of the two-kink bound states in the XXZ spin-1/2 chain model \eqref{XXZH}  in the massive antiferromagnetic phase
in the presence of a staggered magnetic field is  calculated perturbatively in two asymptotic regimes: 
(i) in the extreme anisotropic limit   $\Delta\to-\infty$ to the first order in  $\epsilon=1/|\Delta|$, and 
(ii) for generic $\Delta<-1$ at a weak  magnetic field, to the first order in $h$. Preliminary analysis
shows a good agreement\begin{footnote}
{Note, that due to \eqref{trmod},  only the $\eta=+$ mode with $S^z=0$ 
contributes to the longitudinal structure factor $S^{zz}(\omega,Q=\pi)$,
which is displayed by red curves in Figures 12 - 15 of \cite{Bera17}, 
while the another $S^z=0$ mode with $\eta=-$ contributes 
to the structure factor $S^{zz}(\omega,Q=0)$.}
\end{footnote}
of the obtained analytical representations for the energy spectra with the  results
of numerical  calculations performed by Bera {\it et al.} and presented in Figures 10 - 15 of \cite{Bera17}.
To illustrate this, we have displayed in Figure \ref{Fig:spectra}  by solid curves the magnetic-field dependence
for the lowest mode energy calculated by equation \eqref{lowE} at two values of the anisotropy 
parameter $\Delta=-2$, and $\Delta=-5$. The points in Figure \ref{Fig:spectra} represent the numerical results for the
same energy spectra extracted from  Figure 11 of ref. \cite{Bera17}.  
Taking into account, that no fitting parameters have been used,
the agreement is seen to be excellent. 

Of course, the detailed comparison of the obtained
analytical results with already existing  \cite{Bera17}  numerical and experimental data is required. On the other
hand, it is desirable to validate the obtained results \eqref{lowE} - \eqref{LE54} by reproducing them 
 in a more powerful and systematic approach  based on the 
Bethe-Salpeter equation.

\acknowledgments
I would like to thank Hans  Werner Diehl and  Frank G\"ohmann for fruitful discussions.


\end{document}